\documentclass[usegraphicx, useAMS, usenatbib]{mn2e}

\newcommand{\Msun}{\ifmmode{\rm M_\odot}\else{M$_\odot$}\fi}
\newcommand{\Zsun}{\ifmmode{\rm Z_\odot}\else{Z$_\odot$}\fi}

\title[Velocity dispersion of LCGs at intermediate z]{Stellar velocity
  dispersion of Luminous Compact Galaxies at intermediate redshift}
\author[N. Gruel et al.]
{N. Gruel$^{1,2,3},$\thanks{E-mail: nicolas.gruel@gmail.com}
R. Guzm\'an$^{3}$,
D. Crist{\'o}bal-Hornillos$^{1,2}$,
P. S{\'a}nchez-Bl{\'a}zquez$^{4}$\\
$^{1}$Centro de Estudios de F{\'i}sica del Cosmos de Arag{\'o}n, Plaza San Juan 1, Planta 2, Teruel, 44001,  Spain.\\
$^{2}$Instituto de Astrof{\'i}sica de Andaluc{\'i}a, C/ Camino Bajo de Hu{\'e}tor 50, E-18008 Granada, Spain.\\
$^{3}$University of Florida, 211 Bryant Space Center, Gainesville, FL 32611, USA.\\
$^{4}$Universidad Aut{\'o}noma de Madrid, Ciudad Universitaria de Cantoblanco 28049 Madrid, Spain\\
}

\begin{document}

\date{Accepted 2011 October 19.  Received 2011 October 18; in original form 2011 May 4}

\pagerange{\pageref{firstpage}--\pageref{lastpage}} \pubyear{2011}

\maketitle

\label{firstpage}

\begin{abstract}
  We present the stellar velocity dispersion measurements for 5
  Luminous Compact Galaxies (LCGs) at z=0.5-0.7. These galaxies are
  vigorously forming stars with average SFR $\sim$ 40 M$_{\odot}$/yr.
  We find that their velocity dispersions range from $\sim137\
  \rm{km/s}$ to $260\ \rm{km/s}$, while their stellar masses range
  between $4\times 10^{9}$ and $10^{11}$ M$_{\odot}$.  If these LCGs
  evolve passively after this major burst of star formation, their
  masses and velocity dispersions, as well as their evolved colours
  and luminosities are most consistent with the values characteristic
  of early-type spiral galaxies today.

\end{abstract}


\section{Introduction}

Luminous Compact Galaxies \citep{hammer2001} are starbursts galaxies at
intermediate redshifts mostly detected in both UV and IR wavelengths,
characterised by having small effective radii ($R_e<5\ \rm{kpc}$), high
luminosities ($M_{AB}(B) \leqslant -20$) and strong emission
lines. \citet{hammer2001} observed a representative sample of LCGs selected
from the CFRS fields using the intermediate resolution ($R>600$) spectrograph
FORS1 and FORS2 on the VLT/Kuyen telescope. The spectra revealed some strong
absorption lines (Ca II K and H, G Band, Fe I, and Balmer lines) as well as
narrow and intense emission lines ([OII]$\lambda3727$\AA,
[OIII]$\lambda\lambda4858,5007$\AA\AA, Balmer lines). The spectro-photometric
analysis of these galaxies (\citealt{hammer2001}, \citealt{gruel2002}) showed
that they are likely composed of three different stellar populations. The
youngest population presents strong emission lines ([OII]$\lambda3727$\AA~
Balmer lines), indicating present day active star formation and sub-solar
metallicity. A second stellar population was formed within the last few hundred
million years. It is characterised by a solar metallicity and the presence of
Balmer lines in absorption. The third stellar population, older than 5 Gyr,
exhibits solar metal abundances and strong metallic absorption lines (Calcium,
Iron lines, G band, Titanium, etc.) \citep{gruel2002}.

Almost all LCGs analysed by \citet{hammer2001} have large extinction
coefficients ($A_v\sim 1.5$), yielding average extinction-corrected star
formation rates SFR $\sim 40 \rm{M}_\odot\,\rm{yr}^{-1}$ (which is $\sim$10
times higher than those estimated from the UV fluxes) (\citealt{hammer2001},
\citealt{gruel2002}). Some LCGs show morphological irregularities and/or close
companions as revealed in HST images (\citealt{hammer2001},
\citealt{zheng2004}).  These observations suggest that LCGs may be undergoing
violent star formation events similar to those occurring in close interacting
systems. \citet{hammer2001} concluded that these LCGs may be the progenitors of
the bulges of spirals galaxies forming inside-out.

A new parameter to shed light on LCGs local counterparts is the galaxy's
kinematics. Measurements of the $H\beta$ emission line velocity widths for
Hammer et al. LCG sample showed low velocity widths of $\sim70$ km$^{-1}$,
suggesting they may be low mass objects ($\sim10^9$M$_\odot$).  $H\alpha$
velocity widths for a handful LCGs measured with ISAAC at the VLT
\citep{tresse2002}, showed a ``double horn'' profile characteristic of
rotation.  Integral spectroscopy was also used to measure the velocity field
for some LCGs by \citet{puech2006} using GIRAFFE at the VLT. However, due to
the small apparent size of these objects, the velocity map extends only over
very few spaxels.

The most reliable measurement of galaxy kinematics is the stellar velocity
dispersion from absorption lines. In this paper, we present the first velocity
dispersion measurements from absorption lines for 5 LCGs at intermediate
redshift. We constrain our study to the spectral range that includes the Balmer
lines ($H10, H9, H8, H\epsilon, H\delta, H\gamma$), the calcium doublet
$\rm{Ca}II\ K$,$H$ and the G Band. Section \ref{sec:data} presents the data set
for our galaxy sample.  Section \ref{sec:methods} describes the methods used to
measure the velocity dispersion, the photometry and the stellar masses of these
galaxies. Section \ref{sec:results} contains the results and the discussion.
We assume the following cosmology in this paper: $\textrm{H}_0=70\textrm{ km
  s}^{-1}\textrm{ Mpc}^{-1}$, $\Omega_{\lambda}=0.7$ and $\Omega_{m}=0.3$.

\section{Data}
\label{sec:data}
\subsection{Sample selection}

The galaxy sample was selected from three Canada France Redshift
Survey (CFRS) fields: CFRS 0000+00 \citep{lefevre1995}, CFRS 0300+00
\citep{hammer1995} and CFRS 2230+00 \citep{lilly1995}.  Intermediate-z
LCGs were selected using criteria defined by \citet{hammer2001}: size
($r_{1/2} \leqslant 5\,\textrm{h}^{-1}_{50}\ \textrm{kpc}$),
luminosity ($M_{AB}(B) \leqslant -20.0$), redshift ($0.5 \leqslant z
\leqslant 1$) and the presence of a major star formation episode
characterised by an [OII]$\lambda3727$\AA~emission with an equivalent
width $EW([OII]\lambda3727$\AA$) \geqslant 15$\AA. This results in a
sample of 32 LCGs, or 29\% of the most luminous ($M_{AB}(B) \leqslant
-20.0$) galaxies found in the three CFRS fields (Gruel 2002). We
observed 22 of these galaxies with the FORS/R600 and I600 spectrograph
at the European Southern Observatory 8m VLT/Kueyen at a resolution
of FWHM=5~\AA~(\citealt{hammer2001}, \citealt{gruel2002}).  The
typical exposure times were 12000 seconds per object.

We reduced the spectroscopic data by using the \textit{Multired}
package within \textbf{IRAF}.\footnote{IRAF is distributed by National
  Optical Astronomical Observatory which are operated by the
  Association of Universities for Research in Astronomy, Inc., under
  cooperative agreement with the National Science Foundation.} 
Measurements and analysis (flux, equivalent width, star absorption
correction, metallicity) were done with software described in
\citet{gruel2002}.

In this paper, we analyse a subsample of give LCGs with the higest S/N and the
strongest absorption lines, for which measurements of stellar velocity
dispersion are feasible.

\section{Methods}
\label{sec:methods}

\subsection{Velocity dispersion measurements from the absorption lines}

Velocity dispersion was measured with the program \textit{Movel}, included in
the \textit{REDUCEME} softwares package \citep{cardiel1998}. The instrumental
resolution of our LCG sample is FWHM$\sim5$~\AA~($\sigma_{ins}\sim100\
km/s$). The lowest velocity dispersion measurable with this instrumental
resolution is $\sigma\sim70\ \rm{km/s}$ for spectra with $S/N>10$
(\citealt{matkovic2005}).  We thus selected galaxies from our LCGs sub-sample
with $S/N\ga10$ per resolution element. The S/N criteria reduced the LCGs
sample to only 5 objects. The final sample is given in table \ref{tab:measure}.

Since the original aim of the observations was to analyse emission lines, we
did not observe stars that we could use as templates. Therefore,
we created our own a series of stellar templates with the same instrumental
resolution as the LCGs.

\begin{figure}
\includegraphics[width=8cm]{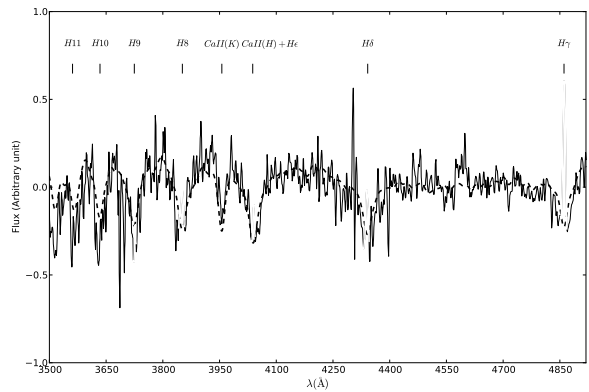}
\caption{Example of one stellar spectrum (22.0637) (plain line) with
  the best broadened stellar template fit (dashed line) obtained with
  \textit{Movel}. The different absorption lines used are labelled.}
\label{fig:spectrum}
\end{figure}

We used stellar templates from MILES spectral stellar library
(\citealt{sanchez-blazquez2006}, \citealt{falcon-barroso2011}), which includes
stars observed between July 2001 and December 2002 in La Palma (Spain) at the
Isaac Newton telescope. These templates are the basis of our galaxy
template. Since their resolution is higher ($\sim2.5$\AA~FWHM) than our sample
of galaxies (R=600), we convolved them with a gaussian kernel. The
characteristics of the gaussians used for the convolution was determined from
the difference between the instrumental resolution of the galaxies and the
stellar database. The instrumental/spectral resolution of the individual LCGs
was determined from the sky spectrum obtained with the same slit as the
spectrum of the galaxy. The intrinsic dispersion of the instrument is given by
$\sigma_{Inst} = \frac{\rm{FWHM_{sky}}}{2.35} \times \frac {1}{\lambda} \times
c$\label{eq:1}. The value of the Gaussian function used to broaden the stellar
template is the quadratic difference of the instrumental resolution of our
observations and that of the stellar library: $\Delta\sigma =
\sqrt{\sigma_{Inst}^2 - \sigma_{stars}^2}$ (See Fig. \ref{fig:spectrum}).

Using a stellar database to build stellar templates by convolving the
instrumental resolution to one's own observations has already been successfully
demonstrated for early-type galaxies by \citet{treu1999}, \citet{treu2001a},
\citet{treu2001b}. The studies proved that at any given S/N and resolution,
there is a lower limit to the velocity dispersion that can be
measured. Typically, this limit stands at half the instrumental resolution of a
given galaxy spectrum at low S/N \citep{bender1992}. All measurements with a
$S/N<10$ per resolution element are deemed not reliable. \citet{treu2001a} and
\citet{matkovic2005} quantified the errors for different S/N and range of
velocity and showed that the systematic error for a galaxy with a S/N$\ga10$
and a velocity dispersion of 150 km/s is $\sim15$\%.

Velocity dispersions for the galaxies were measured using the MOVEL and OPTEMA
algorithms described by Gonzalez (1993). The MOVEL algorithm is an iterative
procedure based in the Fourier Quotient method (Sargent et al.  1977) in which
a galaxy model is processed in parallel to the galaxy spectrum. The main
improvement of the procedure is introduced through the OPTEMA algorithm, which
is able to overcome the typical template mismatch problem by constructing for
each galaxy an optimal template as a linear combination of stellar spectra of
different spectral types and luminosity classes.

For this study, error in velocity dispersion measurement was statistically
determined by bootstrapping. One hundred pseudo-spectra were randomly simulated
in the range given by the galaxy error spectrum using a gaussian noise
model. The error of the velocity dispersion measurement was obtained from the
statistical distribution of the 100 pseudo-spectra measurements.

Different effects limit the precision or the ability to measure galaxy velocity
dispersion. Stellar template mismatch is one such limiting factor. This effect
happens when the star template used to fit the galaxy spectrum are not
representative of the real stellar population of the galaxy. Creating a galaxy
template from a stellar library can be impossible, the stellar library does not
contains every stellar type and one of the main stellar population which
composed the observed galaxy can be missing. As this template is a first guess
to start the velocity dispersion measurement, no numerical criteria to avoid
template mismatch is used. It is possible to help the software to make this
first guess by removing some extra-features (see below) or limiting the number
of stars in the library to the one presenting the most important features for
velocity dispersion measurement. We removed every galaxy spectrum from our
sample when an important template mismatch at the initial step was visually
detected even if their S/N was over $10$. To test the impact of a small initial
template mismatch (no visual detection and $\chi^2$ greater than the template
automatically created by the software) we forced the software to use specific
templates, thus creating an artificial template mismatch. Measurement of the
velocity dispersion with this mismatch showed an error of $\sim5$\% in the
dispersion measurement.

In our case, another effect is introduced by the convolution of the
instrumental resolution of the stellar templates to that of the LCG's. The
effect of this modification was intensively tested with galaxies of well known
velocity dispersion.  An elliptical galaxy from the Coma cluster was changed
and the velocity dispersion measured with the templates convolved to a slightly
different one. The procedure was repeated with a pseudo-instrumental resolution
modified up to $\pm3$\AA~in 0.05\AA~steps. An error of $0.1$\AA~in the
difference of resolution introduces an error of 5\% in the velocity dispersion
measurement. The error in velocity dispersion increases more strongly with
higher error in resolution. At $\pm0.1$\AA~the error is $\sim5$\%, while at
$\pm0.2$\AA~and $\pm0.5$\AA~the errors are respectively $\sim10$\% and
$\sim22$\% in the dispersion measurement. The galaxies instrumental resolution
were measured using different lines from the sky spectrum extracted at the same
time than the LCG spectra with the same polynomial \citep{gruel2002}. The
resolution measured from the different lines was stable at $\pm0.1$\AA~along
the spectrum.

Small emission lines tend to fill the Balmer lines (our major features)
changing the global shape of the absorption features To quantify the influence
of this contamination, the emission lines in our sample were masked and
velocity dispersion was measured without them. We also observed that if not
removed entirely, an artifact can appear at the continuum substraction step for
the strongest emission lines such as [OII]$\lambda3727$\AA~or H$\beta$ and bad
sky substraction features. As a result these lines were erased manually. Both
effects were quantified (measured with and without the lines) as an error of
2\% for the strong emission line $[OII]\lambda3727$\AA~and 5\% for the emission
inside the Balmer lines. For galaxy 03.0645, where the absorption lines were
weaker and contaminated by inside emission, inside emission lines were also
removed and led to an increased error in velocity dispersion.

To emphasise the strongest or the most useful absorption lines, the analysis
has been restrained to the wavelength range
$\lambda\lambda[3570,4400]$\AA\AA. The best absorptions lines present in the
spectra were used (Balmer lines ($H10, H9, H8, H\epsilon, H\delta, H\gamma$),
$\rm{Ca}II\ K$ and $H$ and the G Band). These restriction have no incidence in
the measurement and improve continuum substraction.

As a summary, the error analysis showed that the accuracy of velocity dispersion
measurement is principally limited by the noise in each galaxy spectrum. Our
LCGs sample was thus restrained to galaxies with $S/N>10$. The second limiting
effect is the possible mismatch in resolution between the stellar templates and
the galaxies. This effect was minimised by adjusting the stellar resolution to
the galaxy spectra. Still an error of 10\% was found for a typical error of
0.1\AA.  The average template mismatch was evaluated at around 5\%. Finally
error due to the presence of strong emissions lines, an error purely numerical,
was found to have minimal impact with value of less than 2\%.

\subsection{Photometry}

The B, V, I and K photometry for our sample were obtained from the
CFRS catalogues (\citealt{lilly1995}, \citealt{lefevre1995},
\citealt{hammer1995}).  The absolute AB magnitudes and rest-frame
colours were derived from the best fit to the photometrics data of
models from \citet{bruzual1993} and \citet{bruzual2003}.  The errors
were derived using monte-carlo simulations and estimated to be
$\sim0.14$ and $\sim0.2$ mag for the magnitudes and colours
respectively. Note that all the values were transformed to the
concordance cosmology used in this paper.

\subsection{Stellar mass measurements}

Stellar mass were calculated using the code described in \citet{cristobal2005}
and \citet{hempel2011} based on \citet{guzman2003} idea.  The ages and masses
were determined by fitting the observed photometry (B, V, I and K band) to
the modelled flux obtained from the convolution of a redshifted library of
synthetic galaxy spectra with the filter transmission functions.

A two component synthetic model consisting on a young instantaneous burst and
an exponentially declining SFR was considered. Models were built using BC03
stellar population predictions \citep{bruzual2003} considering different ages
for both components as indicated in Table\,\ref{Tab:model}.  The models depend
on several parameters: IMF, metallicity, $A_V$, extinction law, mass, SFR, and
age.  We considered the same IMF, metallicity, $A_V$ and extinction law for
both old and young component. Therefore a total of 9 parameters (given in
Table\,\ref{Tab:model}) define the two-component population model. To handle
this degenerate problem, we first made simulations (\citealt{cristobal2005b})
to understand the influencing parameters on the stellar mass determination. We
considered a modelled LCBG galaxy with the mean parameters of
\citet{guzman2003} and placed it at $z=0.5$, $z=0.8$, and $z=1.1$, adding
realistic photometric uncertainties (0.03, 0.06, and 0.1 mag for each
redshift). For the simulations we considered the U, B, V, I, K bands. We saw
that extinction and metallicity were the most influencing parameters for final
stellar mass. Increasing the amount of extinction tended to decrease the
inferred stellar mass. Due to the extinction-age degeneracy, a younger and less
massive underlying component increased the extincted flux in the rest-frame UV
bands. Similarly, a lower metallicity also increased the stellar masses. Other
parameters such as SFR, extinction law and IMF had little effect on the
estimates of masses. In our study, the best models ($\chi^2\le \chi^2_{min}+1$)
deviated less than a factor 2 in the inferred stellar masses.

We fixed certain parameters using measurements from the real spectra. The
extinction $A_V$ was measured from the Balmer decrement between H$_\beta$,
H$_\gamma$ and H$_\delta$ emissions lines and infrared flux when available. In
the initial approach we used the Large Magellanic Cloud extinction law for all the
galaxies. For one of them, 03.1540, the models could not fit the SED
correctly. The effect was reduced by using Calzetti law for the extinction law
but still, the fit did not reproduce the expected SED. The mass determined for
this galaxy was at the upper limit and probably overestimated. This LCG was
also detected by ISO and considered peculiar in both term of its extinction and
star formation.

For the fitting process, the ranges considered for the different parameters are
shown in Table\,\ref{Tab:model}.  The $\chi^2$ is calculated from the
difference between observed and modelled photometry as in
equation\,\ref{Fig:eq1}.

\begin{equation}
\chi^2=\sum_{j}\left(\sigma_j\frac{F^{mod}_j-F^{obs}_j}{F^{obs}_j}\right)^2
\label{Fig:eq1}
\end{equation}

where the weights $\sigma_j$ in each filter, are given by $1.086/\sigma_{m_j}$
where $\sigma_{m_j}$ is the error in magnitudes. At each iteration, once the model
parameters are fixed, the fitting process determines which combination of
stellar masses and ages for the components produces the best $\chi^2$ fit to
the observed photometry. A simultaneous fit to the two-component model is
performed.  We attempted an alternative approach, consisting of first fitting
the young burst component to the blue bands photometry, and then fitting the
residuals to the underlying component.  Such approach seemed reasonable since
the bluest/reddest bands were expected to be dominated by the young/old
component.  However, the results were found to strongly depend on which of the
two components was fitted first, i.e. if we first fitted the young component,
then too much NIR flux was assigned to the component, yielding a low total
mass.

The upper age was defined as the age a galaxy formed at $z=4$ would have at the
$z$ observed using our adopted cosmology. We took $10^7$\,yr as the lower age
limit for recent burst of star formation, since the models do not include
nebular component. The lifetime of the stars producing most of the ionising
photons are typically $3-5\times10^6$\,yr; after $10^7$\,yr the production rate
of ionising photons drops by over 99\% \citep{charlot2001}. The simulations
(\citealt{cristobal2005b}) showed that in a two component fitting model, when
the burst is allowed to be younger, the underlying component turns out to be
older and more massive. In that sense the stellar masses estimated here are a
lower limit.

\begin{table}[h]
\begin{center}
\caption{Ranges of the parameters used to model fitting. \label{Tab:model}}
\vspace{0.0cm}
\small{
\begin{tabular}{ll}    
\noalign{\smallskip}
\hline\hline
\noalign{\smallskip}
Parameter	&  Range \\
\noalign{\smallskip}
\hline
\noalign{\smallskip}
Age of recent burst\dotfill & $10^7-5\times10^{8}$\,yr \\
Age of underlying population\dotfill & $2\times10^{8}-2\times10^{10}$\,yr \\
IMF\dotfill & \citep{salpeter1955}\\
Mass of stars: lower, upper limits\dotfill & 0.1, 100\,$\Msun$\\
A$_V$\dotfill & Fixed\\
Extinction Law\dotfill & Fixed\\
Metallicity\dotfill & 0.4 $\Zsun$, $\Zsun$ \\
SFR (underlying)\dotfill & $\tau=1.0$\,Gyr\\
SFR (burst)\dotfill & Instantaneous\\
\noalign{\smallskip}
\hline
\hline
\end{tabular}
}
\normalsize
\rm
\end{center}
\end{table}

The masses measured for the LCGs sample using our own code indicate
that these galaxies have a stellar mass $4.57 \times 10^9\Msun \le
M_\star \le 1.38 \times 10^{11} \Msun$.

To estimate the evolution of these galaxies to $z=0$, the code fits the galaxy
at the observed redshift and we considered the fitted models as they continued
to evolve until $z\sim0$.  We assumed that all gas in the fitted galaxy had
been used to produce stars.

\begin{figure}
\includegraphics[width=8cm]{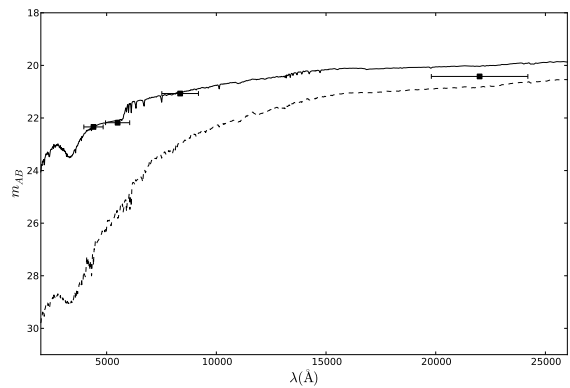}
\caption{Fit of one LCG (22.0637) photometry with two different galaxy
  populations. Plain line, the model is fitting the observed
  photometric points and the dashed line is the model which has
  passively evolved until z=0. The evolved photometric point are
  measured by integrating this spectrum.}

\label{fig:fig2}
\end{figure}

In the models, we are supposing a passive evolution for our LCG. They
are experiencing their last starburst and will not have any mergers. This last
hypothesis is sustained by \citealt{conselice2005} who showed that the major
merger rating decrease since $z\sim1$. Under these conditions, velocity
dispersion and masses are independent of the evolution of the galaxies and the
colour and luminosity will be determined by single evolution of the best fit of
the observed SED. The evolved point used later in this paper was calculated
from the models after evolution (see Fig. \ref{fig:fig2}).

\section{Results and discussion}
\label{sec:results}

The velocity dispersion measurements and the stellar mass estimates 
are listed in the
table \ref{tab:measure}. 
The velocity dispersions from the absorptions lines of the LCGs range
between $\sim137\ \rm{km/s}$ and $260\ \rm{km/s}$ with a median of
$\sim180\ \rm{km/s}$. The two galaxies with highest velocity
dispersion (03.1349 and 03.1540) are ISO galaxies. \citet{hammer1999}
showed that galaxies detected by ISO tend to be large and massive, and
found mostly in interacting systems yielding strong star formation
rates ($>100 \rm{M}_\odot\ \rm{yr}^{-1}$).

\begin{table*}
\caption{Measurements and results}
\begin{tabular} {c|rrrrrrrrr}
name    &  $S/N$   &  $z$  &  $M(B)_{AB}$ & A(V)  &
$[O/H]$&$\Delta[O/H]$   & $\sigma_{abs}$ & $\Delta\sigma_{abs}$ &  $log(M_\star)$\\
        &          &       &  (mag)    & (mag)  &
 &      & (km/s) & (km/s) &  $(M_\odot)$\\
\hline
03.0645 & 12 & 0.527 & -20.45 & 1.53 & 8.57 & $\pm$0.09 & 133.3 & $\pm$91.2 & 10.68\\
03.1349 & 13 & 0.616 & -21.15 & 1.08 & 8.98 & $\pm$0.13 & 196.5 & $\pm$15.9 & 11.14\\
03.1540 & 12 & 0.689 & -21.27 & 3.52 & --- & --- & 259.7 & $\pm$37.3 & 10.89\\
22.0429 & 12 & 0.624 & -20.02 & 2.71 & 8.87 & $\pm$0.15 & 137.0 & $\pm$50.0 & 9.66\\
22.0637 & 14 & 0.542 & -20.95 & 1.20 & 8.67 & $\pm$0.08 & 183.3 & $\pm$10.2 & 9.95\\
\end{tabular}
\label{tab:measure}
\end{table*}

\begin{figure}
\includegraphics[width=9cm]{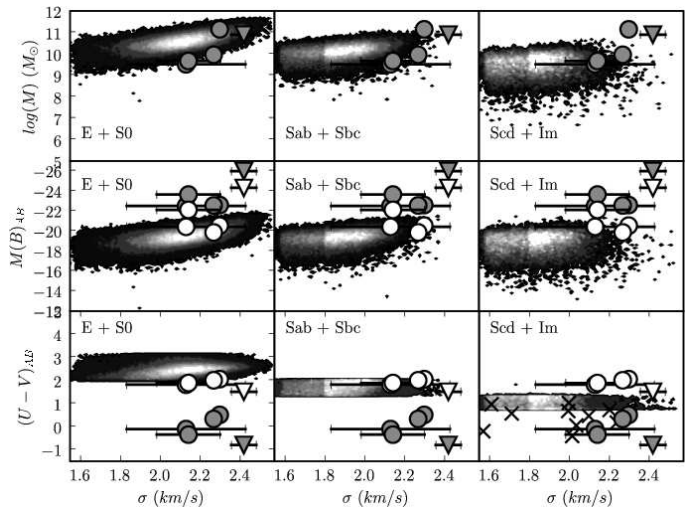}
\caption{In these figures, the points are the LCGs analyse in this
  paper, with a triangle for the ISO galaxy not very well fitted by
  D. Cristobal models. The grey points are the position on the LCGs at
  the time of the observations and in white the same objects after
  passive evolution to $z=0$. To find the nature of the LCGs we
  divided the SDSS data, the reference sample used here, in three
  different categories: Ellipticals, Spirals and Irregulars
  galaxies. The crosses represent a local sample of HII galaxies from
  \citep{telles1997}. The upper row is the $M_{\star}$ versus $\sigma$
  diagram, the middle one the Faber-Jackson relation $L$ versus
  $\sigma$ and the lower the colour versus $\sigma$ relation.}
\label{fig:fig3}
\end{figure}

In Figure \ref{fig:fig3}, we compare stellar masses, $M_B$ and $(U-V)$
colour of LCGs with different type of nearby galaxies.

The data used for comparison include several different kinds of
galaxies taken from the SDSS stellar mass catalogue
\citep{kauffmann2003}.  We remark that this catalogue does have some
degree of mismatch within a $1\arcsec$ radius with the DR4 catalogue we used
for the magnitude and colours. We removed all objects whose cross
correlation yielded different plate identification number, different
fibre identification number, different modified Julian Date,
discrepancies in the redshift or a large error in the measurement of
the mass or the velocity dispersion. Our final selection include
$\sim$94\% of the initial sample.

Figure \ref{fig:fig3} summarise the results of our analysis. The data points
are the 5 LCGs studied in this paper. In grey, the observed value, in white the
model predictions after passive evolution. The triangle corresponds to the
object whose Stellar Energy Distribution (SED) determined by our stellar mass
code did not fit its photometry values. Therefore its estimated evolution is
very uncertain. The crosses in the colour-$\sigma$ diagram are HII galaxies
(see figure caption). The grey scale are the density plots for
different SDSS galaxies. The separation in different galaxies types shown in
the three columns was done by colour selection, following \citet{fukugita1995}.

The upper row is the $M_{\star}$ versus $\sigma$ diagram, the middle
one the $L$ versus $\sigma$ diagram, and the lower the colour versus
$\sigma$ diagram. About $\sim$5 Gyrs ago, the 5 LCGs have stellar
masses and velocity dispersions consistent with those characteristic
of today's population of early-type spheroids and spiral
galaxies. Their luminosities and colours, however, are similar to most
luminous, bluest local HII galaxies. Assuming the stellar masses and
velocity dispersions of these LCGs remain approximately constant, and
that they are undergoing their last burst of star formation, simple
evolutionary synthesis models predict that these objects will evolve
passively to best resemble the typical luminosities and colours of
early-type spiral galaxies. Our velocity dispersion and stellar mass
measurements, combined with this simple evolutionary predictions, are
consistent with the proposed link between LCGs and massive spiral
galaxies \citep{hammer2001}.

We note that our results are not necessarily inconsistent with
previous works suggesting that other class of intermediate redshift 
star-forming galaxies (the so-called Luminous Compact Blue Galaxies)
may evolve into today's population of low mass spheroidal galaxies
(\citealt{koo2005}, \citealt{guzman2003}, \citealt{noeske2006}).  
The LCGs in our sample have a luminosity similar to the 
15\% brightest objects in the LCBGs samples studied by \citet{koo2005}, 
and \citet{noeske2006}. 

\section*{Acknowledgements}

This work is partially funded by the Spanish MICINN under the Consolider-
Ingenio 2010 Program grant CSD2006-00070: First Science with the GTC.
N. G. and DCH acknowledge funding from the Spanish Plan Nacional del Espacio del
Ministerio de Educaci'on y Ciencia (PNAYA2006-14056).  N.G.  and
R. G. acknowledge funding from NASA/STScI grant HST-GO-08678.04-A and LTSA
NAG5-11635. PSB is supported by the Ministerio de Ciencia e Innovaci\'on
(MICINN) of Spain through the Ramon y Cajal programme. PSB also acknowledges a
Marie Curie Intra-European Reintegration grant within the 6th European
framework program and financial support from the Spanish Plan Nacional del
Espacio del Ministerio de Educaci'on y Ciencia (AYA2007-67752-C03-01).  We also
thank the invaluable task of our anonymous referee, which improved greatly the
quality of this manuscript. The authors acknowledge the editorial assistance of
C. M. Perrault, Institut de Bioenginyeria de Catalunya.

\label{lastpage}


\begin{thebibliography}{50}

\bibitem[\protect\citeauthoryear{Bender, Burstein, \& Faber}{1992}]{bender1992}
  Bender R., Burstein D., Faber S.~M., 1992, ApJ, 399, 462

\bibitem[\protect\citeauthoryear{Bruzual A.~\& Charlot}{1993}]{bruzual1993}
  Bruzual A.~G., Charlot S., 1993, ApJ, 405, 538

\bibitem[\protect\citeauthoryear{Bruzual \& Charlot}{2003}]{bruzual2003}
  Bruzual G., Charlot S., 2003, MNRAS, 344, 1000

\bibitem[\protect\citeauthoryear{Cardiel et al.}{1998}]{cardiel1998} Cardiel
  N., Gorgas J., Cenarro J., Gonzalez J.~J., 1998, A\&AS, 127, 597

\bibitem[\protect\citeauthoryear{Calzetti et al.}{2000}]{calzetti2000} Calzetti
  D., Armus L., Bohlin R.~C., Kinney A.~L., Koornneef J., Storchi-Bergmann T.,
  2000, ApJ, 533, 682

\bibitem[\protect\citeauthoryear{Charlot 
\& Longhetti}{2001}]{charlot2001} Charlot S., Longhetti M., 2001, MNRAS, 323, 887 

\bibitem[\protect\citeauthoryear{Conselice, Blackburne, \&
    Papovich}{2005}]{conselice2005} Conselice C.~J., Blackburne J.~A., Papovich
  C., 2005, ApJ, 620, 564

\bibitem[\protect\citeauthoryear{Crist{\'o}bal-Hornillos et
    al.}{2005}]{cristobal2005} Crist{\'o}bal-Hornillos D., Balcells M.,
  Dom{\'{\i}}nguez-Palmero L., Eliche-Moral C., Erwin P., Guzm{\'a}n R., Prieto
  M., 2005, RMxAC, 24, 227


\bibitem[\protect\citeauthoryear{Crist{\'o}bal-Hornillos}{2005b}]{cristobal2005b} 
  Crist{\'o}bal-Hornillos D., 2005, PhDT

\bibitem[\protect\citeauthoryear{Falc{\'o}n-Barroso et 
al.}{2011}]{falcon-barroso2011} Falc{\'o}n-Barroso J., S{\'a}nchez-Bl{\'a}zquez P., Vazdekis A., Ricciardelli E., Cardiel N., Cenarro A.~J., Gorgas J., Peletier R.~F., 2011, A\&A, 532, A95 

\bibitem[\protect\citeauthoryear{Fukugita, Shimasaku, \&
    Ichikawa}{1995}]{fukugita1995} Fukugita M., Shimasaku K., Ichikawa T.,
  1995, PASP, 107, 945

\bibitem[\protect\citeauthoryear{Gonz\'alez-Gonz\'alez}{1993}]{gonzalez-gonzalez1993}
  Gonz\'alez-Gonzalez J.~D.~J., 1993, PhDT

\bibitem[\protect\citeauthoryear{Gruel}{2002}]{gruel2002} Gruel N., 2002, PhDT

\bibitem[\protect\citeauthoryear{Guzm\'an et al.}{1997}]{guzman1997} 
Guzm\'an R., Gallego J., Koo D.~C., Phillips A.~C., Lowenthal J.~D., Faber 
S.~M., Illingworth G.~D., Vogt N.~P., 1997, ApJ, 489, 559 

\bibitem[\protect\citeauthoryear{Guzm{\'a}n et 
al.}{2003}]{guzman2003} Guzm{\'a}n R., {\"O}stlin G., Kunth D., 
Bershady M.~A., Koo D.~C., Pahre M.~A., 2003, ApJ, 586, L45

\bibitem[\protect\citeauthoryear{Hammer et al.}{1995}]{hammer1995} 
Hammer F., Crampton D., Le Fevre O., Lilly S.~J., 1995, ApJ, 455, 88 

\bibitem[\protect\citeauthoryear{Hammer}{1999}]{hammer1999} Hammer 
F., 1999, ASPC, 187, 164 

\bibitem[\protect\citeauthoryear{Hammer}{2000}]{hammer2000} Hammer 
F., 2000, ASPC, 197, 425 

\bibitem[\protect\citeauthoryear{Hammer et al.}{2001}]{hammer2001} 
Hammer F., Gruel N., Thuan T.~X., Flores H., Infante L., 2001, ApJ, 550, 
570 

\bibitem[\protect\citeauthoryear{Hempel et al.}{2011}]{hempel2011} 
Hempel A., Cristóbal-Hornillos, D., Prieto, M., Trujillo, I., Balcells, M., López-Sanjuan,
C., Abreu, D., Eliche-Moral, C., Domínguez-Palmero, L., 2011, arXiv, arXiv:1102.3302 

\bibitem[\protect\citeauthoryear{Kauffmann et 
al.}{2003}]{kauffmann2003} Kauffmann G., et al., 2003, MNRAS, 341, 33 

\bibitem[\protect\citeauthoryear{Kelson et al.}{2000}]{kelson2000} 
Kelson D.~D., Illingworth G.~D., van Dokkum P.~G., Franx M., 2000, ApJ, 
531, 184 

\bibitem[\protect\citeauthoryear{Kobulnicky 
\& Gebhardt}{2000}]{kobulnicky2000} Kobulnicky H.~A., Gebhardt K., 2000, AJ, 119, 1608 

\bibitem[\protect\citeauthoryear{Koo et al.}{2005}]{koo2005} 
Koo D.~C., et al., 2005, ApJS, 157, 175 

\bibitem[\protect\citeauthoryear{Le Fevre et 
al.}{1995}]{lefevre1995} Le Fevre O., Crampton D., Lilly S.~J., 
Hammer F., Tresse L., 1995, ApJ, 455, 60 

\bibitem[\protect\citeauthoryear{Lehnert 
\& Heckman}{1996}]{lehnert1996} Lehnert M.~D., Heckman T.~M., 1996, ApJ, 472, 546 

\bibitem[\protect\citeauthoryear{Lilly et al.}{1995}]{lilly1995} 
Lilly S.~J., Le Fevre O., Crampton D., Hammer F., Tresse L., 1995, ApJ, 
455, 50 

\bibitem[\protect\citeauthoryear{Lilly et al.}{1998}]{lilly1998} 
Lilly S., et al., 1998, ApJ, 500, 75 

\bibitem[\protect\citeauthoryear{Matkovi{\'c} 
\& Guzm{\'a}n}{2005}]{matkovic2005} Matkovi{\'c} A., Guzm{\'a}n R., 2005, MNRAS, 362, 289 

\bibitem[\protect\citeauthoryear{Noeske et al.}{2006}]{noeske2006} 
Noeske K.~G., Koo D.~C., Phillips A.~C., Willmer C.~N.~A., Melbourne J., 
Gil de Paz A., Papaderos P., 2006, ApJ, 640, L143 

\bibitem[\protect\citeauthoryear{Phillips et 
al.}{1997}]{phillips1997} Phillips A.~C., Guzm\'an R., Gallego J., Koo 
D.~C., Lowenthal J.~D., Vogt N.~P., Faber S.~M., Illingworth G.~D., 1997, 
ApJ, 489, 543 

\bibitem[\protect\citeauthoryear{Pisano et al.}{2001}]{pisano2001} 
Pisano D.~J., Kobulnicky H.~A., Guzm{\'a}n R., Gallego J., Bershady M.~A., 
2001, AJ, 122, 1194 

\bibitem[\protect\citeauthoryear{Puech et 
al.}{2006}]{puech2006} Puech M., Hammer F., Flores H., {\"O}stlin G., Marquart T., 2006, A\&A, 455, 119 

\bibitem[\protect\citeauthoryear{Salpeter}{1955}]{salpeter1955} 
Salpeter E.~E., 1955, ApJ, 121, 161 

\bibitem[\protect\citeauthoryear{S{\'a}nchez-Bl{\'a}zquez et 
al.}{2006}]{sanchez-blazquez2006} S{\'a}nchez-Bl{\'a}zquez P., et al., 2006, 
MNRAS, 371, 703 

\bibitem[\protect\citeauthoryear{Sargent et 
al.}{1977}]{sargent1977} Sargent W.~L.~W., Schechter P.~L., 
Boksenberg A., Shortridge K., 1977, ApJ, 212, 326 

\bibitem[\protect\citeauthoryear{Telles 
\& Terlevich}{1997}]{telles1997} Telles E., Terlevich R., 1997, MNRAS, 286, 183 

\bibitem[\protect\citeauthoryear{Tresse et al.}{2002}]{tresse2002} 
Tresse L., Maddox S.~J., Le F{\`e}vre O., Cuby J.-G., 2002, MNRAS, 337, 369 

\bibitem[\protect\citeauthoryear{Treu et al.}{1999}]{treu1999} 
Treu T., Stiavelli M., Casertano S., M{\o}ller P., Bertin G., 1999, MNRAS, 
308, 1037 

\bibitem[\protect\citeauthoryear{Treu et al.}{2001a}]{treu2001a} 
Treu T., Stiavelli M., Bertin G., Casertano S., M{\o}ller P., 2001, MNRAS, 
326, 237 

\bibitem[\protect\citeauthoryear{Treu et al.}{2001b}]{treu2001b} 
Treu T., Stiavelli M., M{\o}ller P., Casertano S., Bertin G., 2001, MNRAS, 
326, 221 

\bibitem[\protect\citeauthoryear{Werk, Jangren, 
\& Salzer}{2004}]{werk2004} Werk J.~K., Jangren A., Salzer J.~J., 2004, ApJ, 617, 1004 

\bibitem[\protect\citeauthoryear{Zheng et 
al.}{2004}]{zheng2004} Zheng X.~Z., Hammer F., Flores H., Ass{\'e}mat F., Pelat D., 2004, A\&A, 421, 847 

\end{thebibliography}
\end{document}